\documentclass[aps,prd,groupedaddress,amsmath,amssymb,amsfonts,nofootinbib,preprint,longbibliography]{revtex4-1}
\usepackage{amssymb,amsfonts}
\usepackage{mathrsfs}
\usepackage{tikz}
\usetikzlibrary{external}
\usepackage{amsthm}
\usepackage{xcolor}
\usepackage{graphicx}
\usepackage[utf8x]{inputenc}
\usepackage{float}
\usepackage{amsmath}
\usepackage{mathtools}
\usepackage{bbold}
\usepackage{accents}
\usepackage{cancel}
\usepackage{esint}
\DeclareMathAlphabet{\mathpzc}{OT1}{pzc}{m}{it}
\usetikzlibrary{patterns}
\usepackage{cleveref}
\pdfoutput=1
\usepackage{epsfig}
\usepackage{graphics}
\begin{document}
\preprint{FERMILAB-PUB-21-329-AE}
\title{Gravity  of Two Photon Decay and its Quantum Coherence}
\author{Kris Mackewicz}
\affiliation{University of Chicago}
\author{Craig Hogan}
\affiliation{University of Chicago}
\affiliation{Fermilab}
\date{\today}
\begin{abstract}
 A linear analytical solution is derived for the gravitational shock wave produced by a  particle of mass $M$ that decays into a pair of null particles.  The resulting space-time is shown to be unperturbed and isotropic,  except for a discontinuous perturbation on a  spherical null shell.   Formulae are derived for the perturbation as a function of polar angle, as measured by an observer at the origin observing clocks on a sphere at distance $R$. The effect of the shock is interpreted physically as an instantaneous displacement in  time  and  velocity when the shock passes the clocks. The  time displacement is shown to be anisotropic,  dominated by a quadrupole harmonic aligned with the particle-decay axis, with  a magnitude $\delta \tau\sim GM/c^3$, independent of $R$. The velocity displacement is isotropic.  The solution is used to derive the  gravitational effect of  a quantum state with a superposition of  a large number of
 randomly oriented, statistically isotropic particle decays. This approach is shown to provide a well-controlled approximation to estimate the magnitude of gravitational fluctuations  in systems composed  of null point particles up to the Planck energy in a causal diamond of duration $\tau= 2R/c$, as well as  quantum-gravitational fluctuations of  black holes and  cosmological horizons. Coherent large-angle quantum distortions of macroscopic geometry from fluctuations up to the Planck scale are shown to grow linearly with the duration, with a variance  $\langle \delta \tau^2\rangle\sim \tau t_P$ much larger than that produced in models without causal quantum coherence.
\end{abstract}
\maketitle
\section{Introduction}

One of the simplest exact solutions of Einstein's equations is the planar gravitational shock wave produced by a point particle on a null trajectory
\cite{Aichelburg1971,DRAY1985173}.
This idealized system has  been applied to study the  ``memory effect'' at null infinity
\cite{Tolish:2014a,Satishchandran:2019}, and extrapolated to quantum systems, including the back-reaction of particle emission on the horizon of a black hole\cite{Hooft:2016itl,Hooft:2016cpw,Hooft2018}. 

Here, we analyze the solution for  the spherical gravitational shock wave produced by a pair of oppositely-propagating point null particles originating from a single point mass at rest in a nearly-flat space-time background. We evaluate the effect of the shock on a sphere at a finite distance $R$, and the observable distortion of  space-time geometry on that sphere as  function of direction and time, as observed at the origin.

In this system, the memory effect takes the form of an anisotropic distortion of time.
 The distortions  can be characterized by a local measurement with a simple  operational definition. Before the particle decays,  a sphere of clocks is synchronized by an outgoing spherical pulse from the origin. After the particle decays,  the gravity of the shock distorts the measured  time on clocks in all directions,  compared synchronously on the world line  of an inertial observer at the origin. 

We choose to study this system in particular because it is well suited to study the relationship of space-time with causally-coherent quantum states of  gravitating mass-energy. 
Unlike the planar shock solution, all of the  elements of the system are contained in a compact causal diamond of space-time.
In this set-up, an entire quantum system, including preparation and measurement of a state and a measurement apparatus,  can be  incorporated into a gravitational solution: both the  preparation of the state  and the measurement of the gravitational perturbation response are  localized on a single world line, in an interval of finite duration. 
In other words, the solution describes a locally-measurable timelike response to  a nonlocal spacelike gravitational effect.

This  solution is applied below to estimate the
effect of causally-coherent quantum nonlocality on gravitational fluctuations.
Our estimate of the  gravitational effect is based on the correspondence principle, so the main conclusions are insensitive to the detailed structure of nonlocal quantum states in curved space-time
 \cite{Pikovski_2017,Banks:2020dus}.

Previous model systems do not have the causal structure required for this exercise.
 The initial conditions for the source in the classical planar shock solution\cite{Aichelburg1971,DRAY1985173}, a null point particle, cannot be set up causally in a  quantum system: a definite direction for the particle momentum requires that its quantum state be delocalized everywhere on planes normal to the direction of travel.  Similarly, the standard approach to nonlocality adopted in effective field theory, widely used to study  gravitational  fluctuations in cosmic inflation\cite{Weinberg:2008zzc,Baumann:2009ds}, quantizes coherent, infinite plane wave modes, not 
 superpositions of directional particle states confined to causal diamonds. 
 
The  angular structure of the spherical null gravitational shock displays a large-angle coherence that is not captured by such models based on nonlocal planar symmetry.
The spherical null shock  relates a causal, nonlocal effect of gravity to localized pointlike events.
The anisotropic, coherent angular gravitational effects in relation to a point are confined to a null surface, and also extend to macroscopic separation. 
The solution is consistent with general relativity and quantum mechanics for particles up to the Planck mass, so it is a suitable approximation to estimate  large-angle correlations of causally-coherent quantum gravity and its fluctuations in macroscopic systems. 

The classical single-particle decay solution is used here to estimate coherent nonlocalized quantum-gravitational effects  of  quantum superpositions of matter states.
We apply the classical solution to two quantum systems, an  $S$-wave decay with an isotropic directional wave function, and the sequential isotropic decay of 
many such particles originating on the same  world line. 
These solutions allow us to derive the large-angle  correlations of  the quantum-gravitational response to a many-particle state.

Assuming that relativity and quantum mechanics are valid and obey the correspondence principle, a superposition of these solutions approximately describes the coherent state of a quantum geometry, up to the point where the total mass of  particles  approaches that of a black hole for the duration defined by the measurement. 
The  extrapolation to many particles provides a controlled estimate of  macroscopic quantum fluctuations of causally-coherent, weak-field gravity for systems of any size. 
The estimate here shows a  variance of large-angle, macroscopic  distortions on the surface of a causal diamond that grows linearly with its duration,  which agrees with recent estimates based on conformal descriptions of near-horizon vacuum states\cite{Banks:2021jwj}, but is much larger than  estimates from  field theory\cite{Weinberg:2008zzc,Baumann:2009ds} that do not include the same causal coherence. This difference is in principle measurable, and might have observable consequences.

\begin{figure}[h]
\begin{centering}
\includegraphics[width=.8\linewidth]{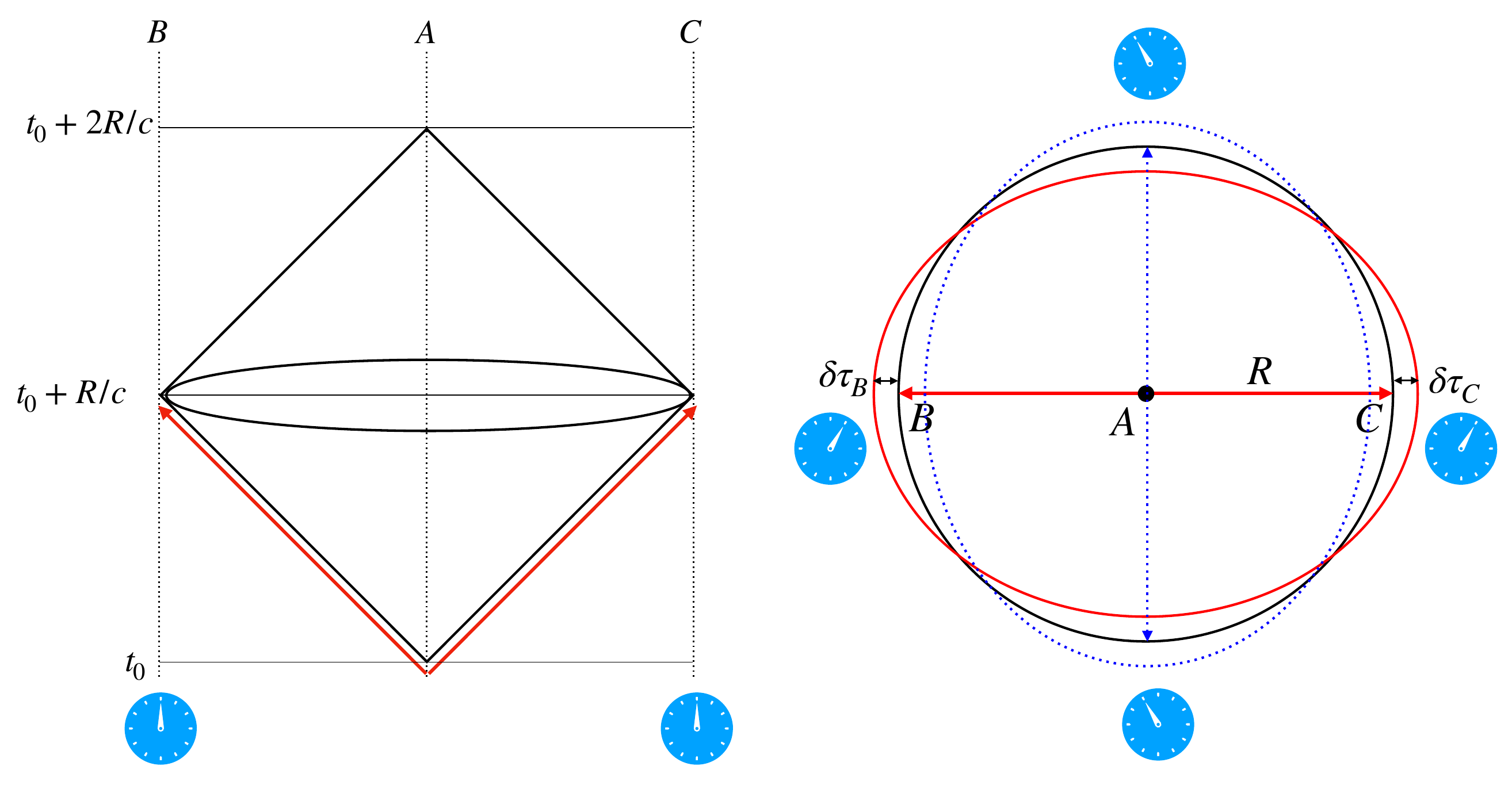}
\par\end{centering}
\protect\caption{Gravitational effect of a particle pair on observed time perturbations. At left, 
a point mass on world line $A$ decays at $t_0$ into oppositely directed null particles.  A time $R/c$ later, the gravitational shock wave of the particles creates a coherent perturbation  on the spherical boundary of a causal diamond
of radius $R$, as shown at the right.  Clocks on the surface are synchronized by an outwards pulse from $A$ just before $t_0$, and observed by $A$  by return light from the surface of the causal diamond after  $t_0+ 2R/c$. The shock creates coherent time displacements with an even-parity directional variation of amplitude $\delta \tau\sim GM/c^3$  aligned with the particle trajectory, as shown by comparing on-axis clocks at $B$ and $C$ with those in equatorial directions. 
 \label{clockdiamond}}
\end{figure}

\section{Setup and general properties of the solution}

Consider the system shown in Fig. (\ref{clockdiamond}). A particle of mass $M$ at the origin is surrounded by a set of clocks distributed on a sphere of radius $R$.  The clocks are synchronized by a pulse from the origin prior to $t_0$. The clocks are on freely falling timelike geodesics. 

At a time $t_0$, the particle decays into a pair of  equal-momentum null point particles that propagate in opposite directions along the $z$ axis. The particles create an outgoing spherical gravitational shock wave that creates a direction-dependent discontinuity in timelike geodesics  as it passes. 
Outside the shock wave, the solution is a  Schwarzschild space-time of mass $M$;  inside, the solution is a flat space-time. 

When the shock passes through $R$, the sphere of clocks is perturbed. The shock creates an instantaneous displacement of position and velocity that depends on the angle from the particle axis.
The bulk of the total  displacement measured as a difference between clocks occurs on large angular scales, that is, in low spherical harmonics, dominated by a quadrupole or tidal term.
The  physical effect of the shock wave  can be visualized  as a coherent,  anisotropic, discontinuous displacement of time 
on the order of $\delta\tau \sim GM/c^3$, together with an isotropic redshift discontinuity $\delta\tau/\tau \sim GM/Rc^2$ that represents the disappearance of gravitational redshift in the post-shock solution.


\section{Spacetime Geometry}
\subsection{Metric Perturbation}
\begin{figure}[ht]
    \centering
    \includegraphics[width=.4\linewidth]{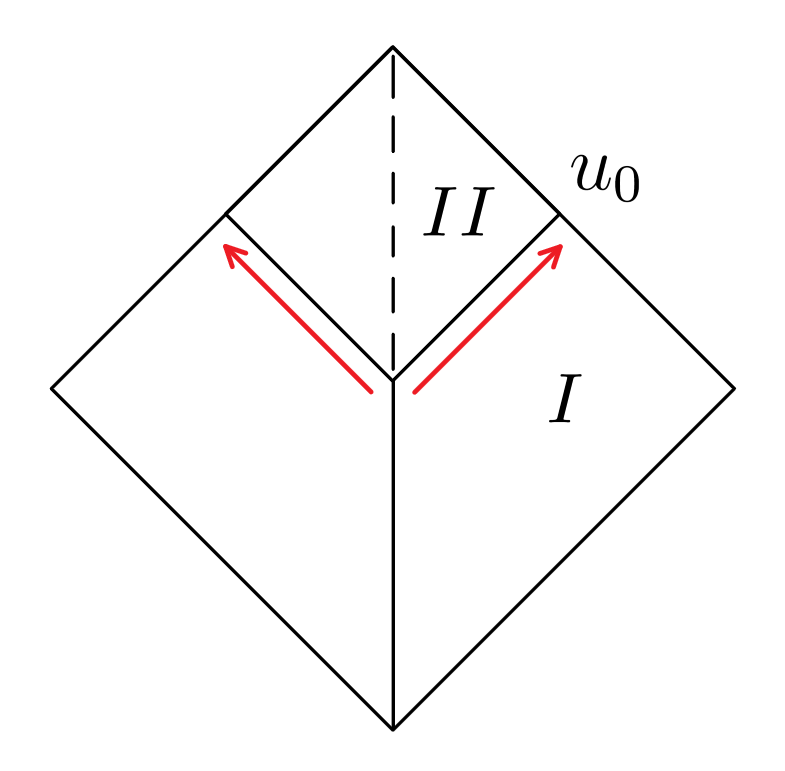}
    \caption{Penrose diagram of decay process. A stationary massive particle sits at the origin (solid world line) and decays at retarded time $u_0$, producing two oppositely propagating null particles. An observer remains at rest at the origin (dashed word line). The null hypersurface created by the spherically propagating shock wave separates the two spacetime regions $I$ (Schwarzschild) and $II$ (Minkowski). }
    \label{fig:decay}
\end{figure}

Consider a particle with mass $M$ sitting at rest at position $z=t_0$ of global inertial coordinates $(t,x,y,z)$. At time $t=t_0$, the particle decays into two massless particles with energy $E=M/2$ that propagate in opposite directions along the $z$ axis. We assume that $M$ is small so that we may use the weak field approximation to determine the spacetime geometry. We will look for the solution to the linearized Einstein equations by considering  a metric perturbation $h_{ab}$ such that $g_{ab}=\eta_{ab}+h_{ab}$. Choosing the Lorenz gauge $\nabla^a \Bar{h}_{ab}=0$, we get a sourced wave equation for the metric perturbation
\begin{equation}\label{waveeq}
    \nabla^c \nabla_c \bar{h}_{a b} = - \frac{16\pi G}{c^4} T_{a b}
\end{equation}
where $\bar{h}_{ab}$ is the trace reversed metric perturbation 
\begin{equation}
    \bar{h}_{ab}=h_{ab}-\frac{1}{2}\eta_{ab} h
\end{equation}
such that $h=\eta^{ab}h_{ab}=-\bar{h}$. (Going forward, we will drop the $G/c^4$ until units are needed). The spacetime diagram for this event is given by \cref{fig:decay}. 

The stress-energy tensor for the system described is
\begin{equation}\label{stressenergy}
    T^{ab}=M\delta(x)\delta(y)\left[\delta(z)H(-t)t^at^b+\frac{1}{2}(\delta(z+t)l^al^b+\delta(z-t)k^ak^b)H(t)\right]
\end{equation}
where $k^a=t^a+z^a, l^a=t^a-z^a$, and $H(t)$ is a Heaviside step function (for simplicity we have set $t_0=0$). For $t<0$, this is the usual stress-energy for a point particle (at rest) given by $T^{ab}=\gamma m v^av^b \delta^{(3)}(\mathbf{x}-\mathbf{x}(\tau))$. For $t>0$, this is the generalization of the previous formula to the massless limit, taking $\gamma m \rightarrow E$.

We can solve this wave equation term by term by letting $\bar{h}^I_{ab} =f_I(x^c) t_a t_b$\,, $\bar{h}^{II}_{ab} =f_{II}(x^c) l_a l_b$\,, $\bar{h}^{III}_{ab} =f_{III}(x^c) k_a k_b$. We then have three sourced scalar wave equations with sources $S_i$ which contain the information of the stress-energy tensor. 
\begin{equation}\label{scalarwaveeq}
    \Box f_i = -4 \pi S_i 
\end{equation}
The retarded Green's function for the $\Box$ operator is 
\begin{equation}
    G_R(t,\vec{x};,t',\vec{x}\,') = \frac{1}{2\pi} \delta(-(t-t')^2+|\vec{x}-\vec{x}\,'|^2)H(t-t')
\end{equation}
The solutions to \cref{scalarwaveeq} are given by
\begin{equation}
    f_i(x) = 4 \pi \int d^4x'G(t,\vec{x};t',\vec{x}\,')S_i(t',\vec{x}\,')
\end{equation}
The complete metric perturbation is then given by \cite{Tolish:2014a}
\begin{equation}\label{metricpert}
    h_{ab}=\frac{2M}{r}(\eta_{a b}+2t_a t_b)H(-U) +\frac{2M}{t+z}l_a l_b H(U) +\frac{2M}{t-z}k_ak_bH(U)
\end{equation}
where $U=t-r$ and $r=(x^2+y^2+z^2)^{1/2}$. The step function behavior indicates that the spacetime will be that of a spherically propagating shock wave. Although at first glance it appears as though the spacetime inside the shock wave ($U>0)$ is not flat, one can make an appropriate coordinate transformation to show that the metric is indeed that of a flat spacetime. We have in null coordinates $u=t-z,v=t+z$
\begin{equation}
    ds^2(U>0) = -dudv + \frac{2M}{u}du^2+\frac{2M}{v}dv^2
\end{equation}
We now define new null coordinates as in \cite{DRAY1985173}
\begin{equation} \label{coord}
    d\tilde{v} = dv - \frac{2M}{u}du \quad d\tilde{u} = du-\frac{2M}{v}dv
\end{equation}
Since we are in the linearized regime, we drop terms $\mathcal{O}(M^2)$. The spacetime metric in the new null coordinates becomes
\begin{equation}
    ds^2(U>0) = -d\tilde{u}d\tilde{v}
\end{equation}
Aside from the discontinuity at $u=0, v=0$ in the new null coordinates $\tilde{u},\tilde{v}$, we see that the spacetime is flat to the causal future of the decay event (henceforth referred to as ``inside'' the shock wave). In the next section we will again demonstrate this by showing that the Riemann tensor vanishes inside the shock wave. 

We briefly mention here that \cref{metricpert} agrees with the metric perturbation derived in \cite{Ratzel:2017} after carefully taking the limit $L\rightarrow 0, \epsilon\rightarrow M/2$, where $L$ corresponds to the width of the laser pulse considered in their work, and $\epsilon$ is the energy of each pulse. Note that while the metric perturbation can me made continuous in their system, the limit $L \rightarrow 0$ makes the metric perturbation have a discontinuous jump at $U=t-r=0$.

\subsection{Geodesic Deviation}
We now determine the relative motion of test bodies in this spacetime (note that in all calculations we raise and lower indices with $\eta_{ab}$). Consider a congruence of timelike geodesics that are initially ``at rest'', with spatial deviation vector $D^a$. Define the tangent to the world line of a test clock by $T^a$. Initially, we have $T^a_0= t^a=(1,0,0,0)$. The deviation evolves according to \cite{Wald:1984}
\begin{equation}\label{geoddev}
    T^e\nabla_e(T^f \nabla_f D^a) = - R_{bcd}^{\quad a}T^b T^d D^c
\end{equation}
Since the clocks will be allowed to free-fall after the system is set up, the tangents to the world lines will pick up $\mathcal{O}(M/r)$ corrections between the time of release and the time of the shock wave passing by. However, we can ignore this correction to the velocity when computing displacement/velocity kicks so long as the free-fall time is not too long since we only care about the leading order effect and the Riemann tensor is at least $\mathcal{O}(M/r)$. The linearized Riemann tensor is given in terms of the metric perturbation by 
\begin{equation}\label{Riemann}
    R_{abcd}=2\nabla_{[a}\nabla_{|[d} h_{c]|b]}
\end{equation}
There will be three types of terms in the Riemann tensor components. One type will be proportional to a step function, which gives the usual gravitational tidal force. The second type will be proportional to a $\delta$ function, which results in a relative velocity kick. The third type will be proportional to the derivative of a $\delta$ function, which results in a relative displacement kick: this is the memory effect. We can write the displacement vector as
\begin{equation}
    D^a=D^a_{(0)}+D^a_{(1)}+D^a_{(2)}+D^a_{(3)}
\end{equation}
where the $(n)$ refers to the $1/r^n$ piece. We consider changes to $D^a$ to linear order in $M$ only. In what follows, we use the following convention for coordinate vectors: $\nabla^a z=z^a,\nabla^a t=-t^a \Rightarrow \nabla^a u=-k^a, \nabla^a v=-l^a$. 

\begin{figure}[h]
    \centering
    \includegraphics[width=.5\linewidth]{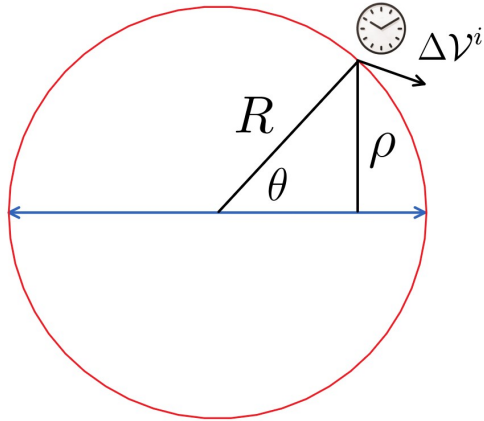}
   \caption{Two oppositely-propagating massless particles (blue) creating a spherically symmetric gravitational shock wave (red). The clocks sitting at radius $R$ experience an instantaneous displacement (not shown) and velocity kick $\Delta\mathcal{V}$ as a function of angle $\theta$. }
    \label{fig:my_label}
\end{figure}

The curvature components proportional to a derivative of a $\delta$ function come from both derivatives acting on the step function. 
\begin{align}
    \nabla_a H(U) &= \delta(U)\nabla_a U \\
    \nabla_d\nabla_a H(U) &= \delta'(U) \nabla_d U \nabla_a U + \delta(U)\nabla_d \nabla_a U \label{ddstep}
\end{align}
Note that since we will ultimately be antisymmeterizing on pairs of indices, we can treat the covariant derivatives as coordinate derivatives in computing the Riemann tensor components. We focus attention on the first term for now. $\nabla_a U = -(t_a+r_a)=-K_a$. Note that at $t=r$, we have $u=r-z=r(1-{\rm cos}\theta), v=r+z=r(1+{\rm cos}\theta)$. Plugging into \cref{Riemann} gives
\begin{align}
    R^{\delta'}_{abcd} &= -\frac{4M}{r}(K_{[a}\eta_{b][c}K_{d]}+2K_{[a}t_{b]}t_{[c}K_{d]})\delta'(t-r) \\ \nonumber 
    &+\frac{4M}{r}\left(\frac{1}{1-{\rm cos}\theta}K_{[a}k_{b]}k_{[c}K_{d]}+\frac{1}{1+{\rm cos}\theta}K_{[a}l_{b]}l_{[c}K_{d]}\right) \delta'(t-r)
\end{align}
If we consider the components of $D^a$ in a parallelly propagated orthonormal frame, the left hand side of \cref{geoddev} reads $d^2D^a/dt^2$. Integrating the geodesic deviation equation twice gives 
\begin{equation} \label{dispkick}
    (\Delta D_{(1)})_a = \frac{M}{r}(\theta_a\theta_b-\phi_a\phi_b)D^b_{(0)}
\end{equation}
where $\theta_a,\phi_a$ are the unit tangent vectors on the sphere. 
Thus, there is an instantaneous relative displacement kick between two nearby stationary test bodies. 

As shown in the appendix, we can decompose this tensor into a tensor harmonic mode sum. Specifically, the displacement kick can be decomposed into a sum over transverse ``electric'' tensor harmonics. The coefficients of the mode sum are given by \cref{tecoeff}:
\begin{equation}
    \Delta D =- \frac{M}{r}(\Hat{\theta}\Hat{\theta}-\Hat{\phi}\Hat{\phi})\cdotp D_{(0)}\sum_{l=2,{\rm even}}^{\infty} \frac{2l+1}{l(l-1)(l+1)(l+2)}\left(2 {\rm cot}\theta \frac{d}{d\theta}+l(l+1)\right) P_l({\rm cos}\theta)
\end{equation}
where $\Hat{\phi},\Hat{\theta}$ are angular unit vectors on the sphere. The quadrupolar mode ($l=2$) gives
\begin{equation}
    \Delta D_{l=2} = \frac{5M}{8r}{\rm sin}^2\theta (\Hat{\theta}\Hat{\theta}-\Hat{\phi}\Hat{\phi})\cdotp D_{(0)}
\end{equation}

We note here that one could attempt to ``glue'' the two spacetime regions (Minkowski and Schwarschild) together using the formalism developed by Israel and Barrabes \cite{Israel:1991}. The results of their work indicate that the Riemann tensor is at most $\delta$ function singular on the null boundary when the induced metric on the null hypersurace joining the two spacetimes can be made continuous. However, as shown in \cite{Satishchandran:2019}, there must be a change in the leading order metric (which is associated with memory) for the type of stress-energy considered here. Therefore, the metric cannot be made continuous here and there will be a $\delta'$ type singular behavior on the null hypersurface.

The curvature components proportional to a $\delta$ function come from the second term in \cref{ddstep} and the cross term when acting two derivatives on $h^I_{ab}$ in \cref{metricpert}. Again, derivatives acting on $1/u,1/v$ will not contribute since they are proportional to $k^a,l^a$ respectively. In addition to the terms mentioned above, there will be a term proportional to $\delta(t-r)$ that comes from integration by parts of the terms proportional to $\delta'(t-r)$ when integrating the geodesic deviation equation: 
\begin{align}
    R^{\delta}_{abcd} &= \frac{4M}{r^2}( \theta_{[a}(\eta_{b][c}+2t_{b]}t_{[c})\theta_{d]}+ \phi_{[a}(\eta_{b][c}+2t_{b]}t_{[c})\phi_{d]}) \\ \nonumber
    & - \frac{4M}{r^2}(K_{[a}(\eta_{b][c}+2t_{b]}t_{[c})r_{d]}+r_{[a}(\eta_{b][c}+2t_{b]}t_{[c})K_{d]}) \\ \nonumber
    & - \frac{4M}{r^2(1-{\rm cos}\theta)} (\theta_{[a}k_{b]}k_{[c}\theta_{d]}+\phi_{[a}k_{b]}k_{[c}\phi_{d]}) \\ \nonumber
    & - \frac{4M}{r^2(1+{\rm cos}\theta)} (\theta_{[a}l_{b]}l_{[c}\theta_{d]}+\phi_{[a}l_{b]}l_{[c}\phi_{d]}) \\ \nonumber
    &+\frac{4M}{r^2(1-{\rm cos}\theta)^2}K_{[a}k_{b]}k_{[c}K_{d]} +\frac{4M}{r^2(1+{\rm cos}\theta)^2}K_{[a}l_{b]}l_{[c}K_{d]}
\end{align}
Again, keeping only the leading order behavior in $1/r$ and integrating the geodesic deviation equation once then gives
\begin{equation} \label{relvelkick}
    \Delta v^{(2)}_a = \frac{M}{r^2}\left[r_ar_b+(\theta_a\theta_b-\phi_a\phi_b)\frac{1+{\rm cos}^2\theta}{1-{\rm cos}^2\theta}+\frac{2{\rm sin}\theta {\rm cos}\theta}{1-{\rm cos}^2\theta}(r_a\theta_b+r_b\theta_a)\right]D^b_{(0)}.
\end{equation}
Thus, there is an instantaneous relative velocity kick between two nearby stationary test bodies as measured in a parallely propagated frame. 

The curvature components proportional to a step function come from both derivatives hitting the pre-factors in \cref{metricpert}. Only the first piece of $h_{ab}$ will contribute since $\nabla^a u = -k^a$,  etc., and the antisymmetrization will kill these terms. The Riemann tensor associated with tidal forces is
\begin{align}
    R^{\rm tid}_{abcd} = \frac{4M}{r^3}(\eta_{[a[d}+2t_{[a}t_{[d})(\delta_{b]c]}-3r_{b]}r_{c]}) H (-(t-r))
\end{align}
Finally, \cref{geoddev} gives 
\begin{equation}
    \frac{d^2 D^{(3)}_a}{dt^2} = \frac{M}{r^3}(2 r_a r_b-\theta_a\theta_b-\phi_a\phi_b)D^b_{(0)} H (-(t-r))
\end{equation}
This is the usual form of tidal forces in Newtonian gravity. Note that this term ``turns off'' after the shock has passed a given test body.

\section{time displacement}
Consider a system of synchronized clocks distributed on a sphere of radius $r$. After the passage of the gravitational shock wave, the clocks will no longer be synchronized due to the relative displacement kick. The desynchronization of clocks on the sphere can be determined via supertranslations at future null infinity, or $\mathcal{I}^+$ \cite{Strominger:2014}. These are asymptotic symmetries (part of the BMS group) of asymptotically flat spacetimes \cite{Bondi:1962}\cite{Sachs:1962}. The generator of these symmetries is given by
\begin{equation}
    \psi^a = T(x^A)\left(\frac{\partial}{\partial u}\right)^a-T(x^A)\left(\frac{\partial}{\partial r}\right)^a -\frac{1}{r}\mathcal{D}^C T(x^A) \left(\frac{\partial}{\partial x^C}\right)^a +...
\end{equation}
which creates an infinitesimal shift in the retarded time
\begin{equation}
    u \rightarrow u - T(\theta,\phi)
\end{equation}
Here $x^A$ are the position coordinates on the 2-sphere. In 4 spacetime dimensions, it has been shown that (null) memory is purely of scalar type, and can be written in terms of these supertranslations \cite{Satishchandran:2019}. The relation is given by
\begin{equation}
    \Delta_{AB} = \frac{1}{r}\left(\mathcal{D}_A\mathcal{D}_B-\frac{1}{2}q_{AB}\mathcal{D}^2 \right) T(\theta,\phi)
\end{equation}
where $\mathcal{D}_A$ is the covariant angular derivative operator on the sphere \cite{Hollands:2016}. From eq. (18) we have
\begin{equation}
    \Delta_{\theta \theta} = \frac{2M}{r}\sum_{l=2,{\rm even}}^{\infty} \frac{2l+1}{l(l-1)(l+1)(l+2)}\left(\frac{d^2}{d\theta^2}+\frac{1}{2}l(l+1) \right) P_l ({\rm cos}\theta)
\end{equation}
From this one can easily read off the $l>1$ contributions to $T(\theta,\phi)$:
\begin{equation}\label{timedisplace}
    T(\theta,\phi)= T_{0,1}+2M\sum_{l=2,{\rm even}}^{\infty} \frac{2l+1}{l(l-1)(l+1)(l+2)} P_l ({\rm cos}\theta)
\end{equation}
The $l=0,1$ contributions are due to standard temporal and radial spatial translations. By setting up our system of clocks on the sphere to be initially synchronized, these two terms vanish. 

Equation (\ref{timedisplace}) is the main result of this paper, the  angular pattern of time displacement recorded on the surface of the causal diamond.
Since the weighting falls off as $\sim 1/l^3$, the $l=2$ mode will be the dominant contribution. It has an angular dependence
\begin{equation}\label{ell2pattern}
    T_2(\theta) = \frac{5M}{24}(3 {\rm cos}^2\theta-1).
\end{equation}

%

\section{Velocity Kick}


Previously, we considered the relative velocity induced between nearby observers in a parallely propagated frame. However, we would also like to know the (global) radial velocity (away form the origin) of each clock on the sphere, as this will produce a longitudinal doppler shift as measured by the observer at the origin. We can determine the induced velocity $\mathcal{V}^a = dx^a/dt$ of each clock by solving the geodesic equation.
\begin{equation}\label{geodesic}
    \frac{d \mathcal{V}^{a}}{dt}+\Gamma^{a}_{b b}\mathcal{V}^{b} \mathcal{V}^{c} = 0
\end{equation}
where we have chosen to parameterize the wordlines by the coordinate time $t$. The tangent vector to the worldline of the clocks is initially given by that of a stationary observer, i.e. $\mathcal{V}^a_0=t^a$. Since the leading order corrections to $\mathcal{V}^a$ are $\mathcal{O}(M/r)$ and the linearized Christoffel symbols are also $\mathcal{O}(M/r)$, we can use $\mathcal{V}^a_0$ in \cref{geodesic}. Therefore in our linearized regime we have
\begin{equation}
    \frac{d\mathcal{V}^i}{dt} \approx - \Gamma^i_{tt}
\end{equation}
where 
\begin{equation}
    \Gamma^i_{tt} = \frac{1}{2}(2\partial_t h_{it}-\partial_ih_{tt})
\end{equation}

We want to determine the instantaneous (globally measured) velocity kick, which comes from the Christoffel symbols proportional to delta functions. In Cartesian coordinates, we have
\begin{equation} 
     \mathcal{V}^{\mu} = \mathcal{V}_0^{\mu} +\frac{M}{x^2+y^2}\left(\frac{r^2+z^2}{r^2},-x \frac{r^2+z^2}{r^2},-y\frac{r^2+z^2}{r^2},z\frac{3r^2-z^2}{r^2} \right) 
\end{equation}

At this point we must discuss a subtle point about what the observer at the origin is measuring. As mentioned in section $IIIA$, the original coordinate chart $(t,x,y,z)$ is not well suited to describe the flat spacetime that the observer at the origin is making measurements in. We want to know how fast a given clock is moving away from the origin as measured in time $\tilde{t}$. We need to make the appropriate coordinate transformation to the coordinates $(\tilde{t},\tilde{x},\tilde{y},\tilde{z})$ and re-parameterize the wordline of the clocks.
\begin{equation}
    \tilde{\mathcal{V}}^a = \frac{dt}{d\tilde{t}} \frac{\partial \tilde{x}^a}{\partial x^b} \mathcal{V}^b
\end{equation}
Using \cref{coord} we find that
\begin{equation}
     \frac{\partial \tilde{x}^a}{\partial x^b} = \delta^a_{b} + \left(\frac{M}{u}+\frac{M}{v}\right)\left(z^a z_b - t^a t_b \right)+ \left(\frac{M}{v}-\frac{M}{u}\right)\left(t^a z_b - z^a t_b \right)
\end{equation}
\begin{align}
    \frac{d \tilde{t}}{dt} = \mathcal{V}^{a} \nabla_{a} \tilde{t} = \mathcal{V}^a\left(t_a\left(-1+\frac{M}{u}+\frac{M}{v}\right)+z_a\left(\frac{M}{u}-\frac{M}{v}\right)\right)
\end{align}
In what follows, we expand everything out to $\mathcal{O}(M/r)$. In the new coordinates $(\tilde{t},\tilde{x},\tilde{y},\tilde{z})$ we have
\begin{equation} \label{velcomp}
     \tilde{\mathcal{V}}^{\mu} = \left(1,-\frac{M}{r} \frac{1+{\rm cos}^2\theta}{1-{\rm cos}^2\theta} {\rm sin}\theta {\rm cos}\phi,-\frac{M}{r} \frac{1+{\rm cos}^2\theta}{1-{\rm cos}^2\theta} {\rm sin}\theta {\rm sin}\phi,\frac{M}{r}{\rm cos}\theta \right) 
\end{equation}

We note here that although we should have transformed $(r,\theta)$ into $(\tilde{r},\tilde{\theta})$, the difference between these coordinates is $\mathcal{O}(M/r)$, so it is sufficient to use the original spherical polar coordinates to leading order (this is explained in more detail in the next section). As can be seen by \cref{velcomp}, there is an inward relative velocity transverse to the $\phi$ direction, with an outward velocity in the $z$ direction. This distorts the sphere into an ellipsoidal shape. If one does a Taylor expansion about a particular point, they will recover the relative velocity kick between nearby observers derived in \cref{relvelkick}.  

Projecting the velocity kick into the radial and transverse directions, we get
\begin{equation}\label{exactkick}
    \Delta \mathcal{V}^r = -\frac{M}{r}
\end{equation}
\begin{equation}
    \Delta \mathcal{V}^{\theta} = -\frac{2M}{r}{\rm cot}\theta
\end{equation}
We find that the radial velocity kick is isotropic, while the transverse velocity kick is not isotropic and is badly divergent near the polar axes, which we discuss how to handle in the next section. This isotropic radial velocity kick can be decomposed into two parts, as demonstrated in \cite{Ratzel:2017}: an outward impulse purely due to the loss of mass in the spacetime, and an inward impulse due to the propagating photon, whose gravity always points inwards and transverse to the $z$ axis (i.e. in the $-\Hat{\rho} $ direction). The net radial impulse is toward the origin, and since it is independent of polar angle, only the angular velocity kick ``knows'' about the location of the counter-propagating null particles - one can show that the relative kick in this direction given by the 2nd term in \cref{relvelkick} agrees with the result of \cite{Aichelburg1971} in the appropriate limit. Although the angular component is divergent at the poles, one can expand for small ${\rm cos}\theta$ (i.e. near the equator) to get an approximate solution. We comment here that the full solution to the non-linear Einstein equations (with smoothed source) should be regular at the poles and should match on to this expansion around the equator in the weak field limit. 

The effect of the radial velocity kick is to produce a Doppler shift in the clocks relative to the central observer.  All the clocks  appear to run faster after the shock has passed; this can be interpreted as an inward velocity kick from the shock, which is indistinguishable from the disappearance of the isotropic  gravitational redshift relative  to the central observer that was present in the pre-shock system. The longitudinal Doppler shift is first order in the (radial) velocity, but the profile is spherically symmetric and thus the observer gains no information about which way the two particles went. If one considers many such decays originating from the same point, there will be no fluctuations in this measurement. The transverse Doppler shift at leading order is second order in the (angular) velocity kick, so this will be a sub-leading effect (recall that $M/r<<1$ to satisfy the linearized regime).

We  conclude that the anisotropic effects at leading order in $M/r$ are due to the time shift induced by the relative displacement kicks of the clocks, which  give rise to measurable angular fluctuations.

\section{ Limits of Validity}

This solution is not exact.  The  linear approximation to gravity breaks down at
large $M$, and because of quantum uncertainty,   the point-particle approximation breaks down at small $M$. (We will continue to assume that the clocks are represented classically, as point-like tracers of local proper time; this does not affect our conclusions in the regime of interest.)

\begin{figure}
\begin{centering}
\includegraphics[width=0.7\linewidth]{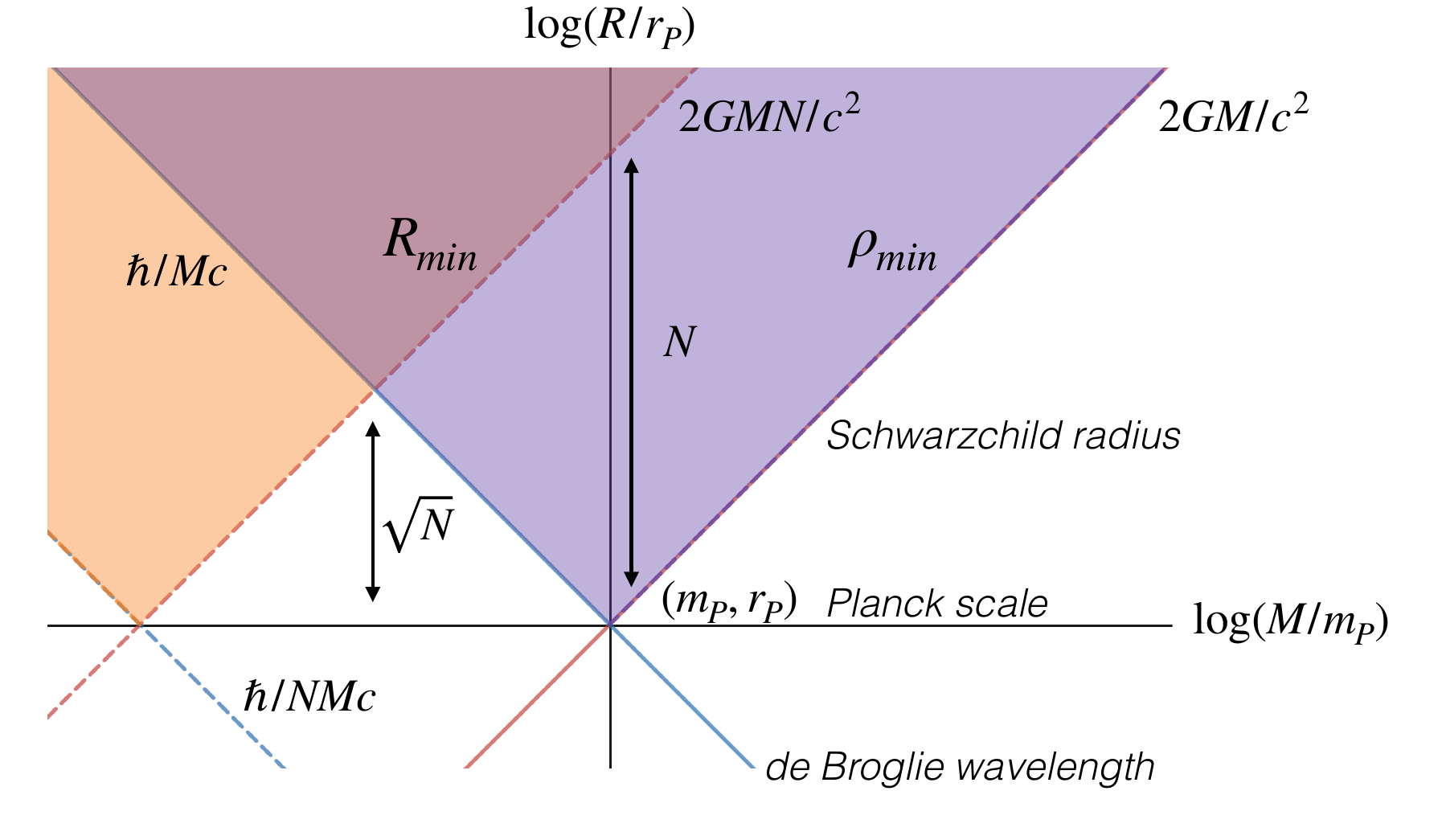}
\par\end{centering}
\protect\caption{Shaded regions schematically show the range of validity for the spherical decay solution imposed by linearity of classical gravitational distortion at large $M$, and localization of quantum mass-energy of the decaying particle at small $M$. For a system of $N$ particles, the boundaries are displaced  as shown.
 \label{validity}}
\end{figure}

For a single  particle decay, the linear approximation  breaks down 
close to the axis.  It leads to unphysical effects if  the impact parameter is less than the Schwarzchild radius for mass $M$: 
 the scattering angle becomes large, and the orbits of clocks carry them across the singularity on the axis.
Putting back the units, from \cref{metricpert} we see that the linear approximation is valid only for 
\begin{equation}
    \frac{2GM}{c^2(t\pm z)} << 1
\end{equation}
and at angles from the axis greater than
\begin{equation} \label{thetalim}
   \theta^2 \gtrsim \frac{4GM}{R c^2}.
\end{equation}
Thus, for any $M$ there is also a lower bound on $R$
(see Fig. \ref{validity}). 

Upon making the coordinate transformation given by \cref{coord}, angles on the sphere will become distorted, i.e. $\theta\neq \tilde{\theta}$. However, one can show that the leading order difference satisfies
\begin{equation}
    {\rm cos}\tilde{\theta}-{\rm cos}\theta \sim \frac{GM}{Rc^2} {\rm ln}\left(\frac{1+{\rm cos}\theta}{1-{\rm cos}\theta}\right)
\end{equation}
So long as $R$ is much larger than the Schwarzschild radius and we are not too close to the poles, we may use $\theta \approx \tilde{\theta}$ in all of our formulas since the displacement and velocity kicks are $\mathcal{O}(M/r)$, and the difference in angles will contribute at sub-leading order. 


For many particles, the solution also requires $R$ to be much larger than the Schwarzchild radius for the whole central mass.  
For $N$ particles of mass $M$ on the same world line, we require
\begin{equation}
R\gtrsim R_{min} = 2GNM/c^2;
\end{equation}
otherwise, the sphere of clocks is inside the Schwarzschild radius of the mass, forming a black hole.



So far, we have considered only the limits of the  classical solution.
In a real physical system, the quantum properties of a  point particle  add other constraints.
The mass-energy of the particle is delocalized in space and time by an amount that depends on the character of its quantum state.
For example,  the de Broglie wavelength,  $\lambda_M=\hbar/Mc$,  gives the minimum spatial uncertainty of position for a massive particle that is localized in time.
This value is shown in Fig. (\ref{validity}).

In the classical solution, the source is a single pointlike object with a definite trajectory. In quantum reality, it is a single quantum object whose wave function in space  depends on its localization in time. 
Temporal coherence leads to  correlations that are nonlocal in proper time, which affects the coherence of the metric distortion in the timelike direction.


The wave packet of particle  location spreads out in space and  time.
The minimum spatial  width  of a location wave function  for a measurement of duration  $\tau$, or ``standard quantum uncertainty''\cite{Caves1980,Caves1980a}, is
 given by
\begin{equation}
\langle\Delta x^2\rangle > \hbar \tau/NM
\end{equation}
for  a state of total mass $NM$.
On the surface of a causal diamond with $R= c\tau$,
the spatial uncertainty in the central world line leads to an angular quantum uncertainty,
\begin{equation}
\langle\Delta \Theta^2\rangle \sim 
\langle\Delta x^2\rangle/R^2\  \sim  (c\tau/R)(\hbar/NMc) R^{-1}\ >\ \hbar/NMcR.
\end{equation}
Thus, the quantum bounds become important when both $M$ and $R$ are small.
Fig. (\ref{validity}) shows the range of classical and quantum validity  in Planck units:
\begin{equation}
m_P\equiv \sqrt{\hbar c/G}, \ \ r_P\equiv ct_P\equiv  \sqrt{\hbar G/ c^3}=\hbar/m_Pc.
\end{equation}



\section{Quantum superpositions}

\subsection{Quantum superposition of randomly oriented decays}

In the region of validity shown in Fig. (\ref{validity}), the calculation far from the polar axis is a good approximation for both general relativity and quantum mechanics. In that regime, it  can be reliably extrapolated to consider quantum states of the nonlocalized metric of a causal diamond. The entire setup, including the geometry itself and the sphere of clocks used to measure it, can be treated as a coherent quantum system.

If a decaying particle is a quantum system,  the quantum nonlocality of the  ``spooky''  two-particle decay state also leads to a delocalized, Schr\"odinger-cat-like macroscopic superposition of space-times. 
As usual in quantum mechanics, there is no paradox: if the preparation of  states is causal, the
quantum coherence of the decaying-particle products entangles with the coherent causal diamonds of the distorted space-time, and the corresponding displacements of clocks.

 Suppose  the null particles from a decay are created in a quantum $S$-wave state, whose wave function is an isotropic superposition of all directions.
This leads to a  superposition of space-time distortion patterns.  The wave function of the space-time includes a superposition of all the different orientations of the classical gravitational  distortions,  each one an eigenstate of the particle decay axis. Each metric in the superposition is coherent over a macroscopic causal diamond, starting with the decay and ending with the null incoming reflection from the boundary that carries the clock readings back to the observer.    The wave function can be described as a sum of state amplitudes over histories, where the sum includes not just a world line, but a whole causal diamond.  A measurement of a particle axis will always find itself with a causal-diamond history consistent with it.


The  classical solution shows that fractional distortion $\Delta$ of time displacement, as measured by observed clocks,  varies coherently on  the causal diamond surface of radius $R$ according to Eq.(\ref{ell2pattern}),
\begin{equation}
\Delta=\delta \tau/(R/c)=  c T_2(\theta)/R \simeq \frac{5Mc}{24R}(3 {\rm cos}^2\theta-1)
\end{equation}

The system with a single particle  can  be extrapolated by linear superposition to a  system of many particles. 
 If  there are $N$  decaying particles  of mass $M_i$ in the causal diamond,
the large-angle coherence does not decrease from averaging their effects: instead, the amplitudes of their distortions linearly add. 
The coherent quantum state of time distortions on the sphere from $N$ such decays on the same world line can be written as a sum
\begin{equation}\label{clockstate}
    | \Delta (\theta, \phi)\rangle = \sum_{i=1}^N \alpha_i (\vec n_i) | \Delta (\theta, \phi)_i\rangle,
\end{equation}
where each element of the sum represents  a decay along a different random axis $\vec n_i$.

Each decay creates the same universal distortion pattern relative to its own axis. Their sum does not give a universal pattern, but it does give  a universal power spectrum:
the spherical harmonic distortion coefficients $\alpha_{\ell m}$ from each decay pattern
$\Delta (\theta, \phi)_i$ add in quadrature.
For example, for equal mass particles $M=M_i$, the  quadrupolar variances add  to give fluctuations with a total quadrupolar variance
\begin{equation}\label{tauvariance}
\langle\Delta^2\rangle_{2} = \sum_{m=-2}^{2}\sum_{i=1}^N  \alpha_{2m}(i)^2  \sim  N  (GM/R c^2)^2,
\end{equation}
where  $\alpha_{2m}(i)$ denote the  quadrupolar amplitudes for each decay, derived above (Eq. \ref{ell2pattern}).
Note that for a given total mass $NM$, the  gravitational distortion decreases with $N$ but increases with an adopted granularity scale $M$.
This result depends only on standard physics, and a conservative application of the quantum correspondence principle, that the gravity of a quantum system is the same as that of an identical classical one.

The same  estimate applies to   large-angle gravitational distortions created by noninteracting massless null particles  that enter and leave the boundary of a volume with radius $R$.
For a  gas with 
with a mean occupation of $N$ particles inside $R$, Eq. (\ref{tauvariance}) approximately gives the variance of  large-angle gravitational fluctuations on the surface.

 This framework is self-consistent until  $N$  is large enough that the gravity of the particles affects the mean curvature of the causal diamond. 
For consistency, in order not to form a black hole with radius less than $R$, we require
\begin{equation}
R< R_S = 2G (N M)/c^2
\end{equation}
so
their number must
not exceed
\begin{equation}\label{Nmax}
N_{max}\sim c^2R/2GM. 
\end{equation}
For a gas of particles that saturates the bound $N=N_{max}$, the   quadrupolar gravitational redshift distortion,
\begin{equation}\label{fractionalvariance}
\langle\Delta ^2\rangle_2 \sim   GM /2R c^2,
\end{equation}
decreases with  $R$, which shows that a large system has a nearly-determinate, classical metric.
Note that Planck's constant $\hbar$ does not appear in Eq.  (\ref{fractionalvariance}): although it is a quantum uncertainty, based on superposition of multiple decays, its magnitude is determined by the assumed discreteness scale, $M$. 


\subsection{Coherent quantum-gravitational fluctuations}

For particle states confined to a volume of size $R$,  extrapolation to $N_{max}$ in Eq. (\ref{fractionalvariance}) gives an estimate of  the gravitational fluctuations created by  randomly-oriented  particle states  with a UV cutoff $M$.  
This  applies approximately to the case of a  system  filled  with  particles whose density approximately saturates the bound, such as a cosmological solution or black hole on the scale of the horizon. For a UV cutoff at the  Planck scale  $M= m_P$,  Eq. (\ref{fractionalvariance})  with $R=c\tau$ becomes
\begin{equation}\label{Deltamag}
\langle\Delta^2\rangle_2\sim \langle\delta \tau^2\rangle/\tau^2\sim  t_P/\tau.
\end{equation}

As expected, quantum-gravitational fluctuations are of the order of unity for $\tau$ at the Planck scale.
The surprising feature of Eq. (\ref{Deltamag}) is the linear inverse dependence on $\tau$ for durations $\tau$ much longer than $t_P$.
The gravitational effects of vacuum field fluctuations in effective field theory have long been studied in the context of  inflationary universes (e.g., ref.\cite{STAROBINSKY1982175,Weinberg:2008zzc,Baumann:2009ds}).
In these systems,  gravitational quantum  uncertainty   decreases as a higher power with scale:   the typical relic metric fluctuation on the scale of an inflationary horizon of radius $c\tau$ is
 \begin{equation}
     \langle\Delta^2\rangle_{EFT}\sim (t_P/\tau)^2.
 \end{equation}
 
 The different results arise from  different  models of quantum coherence. 
The larger fluctuations in the decaying-particle system can be traced to the fact that a point particle creates a  displacement that is coherent on scale $c\tau$ for a causal diamond of any duration $\tau$. The  coherence scale in the effective field system is that of the particle wave function, $\sim  \hbar/Mc$,  so  the  variance  is reduced  by a factor $\sim 1/\tau$.
Our classical solution explicitly shows that  the physical gravitational effect  corresponds to the former case: the gravity of a point particle imprints a large-angle, quasi-tidal coherent geometrical structure on macroscopic scales, much larger than the de Broglie wavelength of the particle.

For  causally-coherent gravity, the large-angle, macroscopic distortion actually increases with  duration, with variance 
\begin{equation}\label{distortionvariance}
 \langle\delta \tau^2\rangle_2\sim  \tau \ t_P.
\end{equation}
This scaling behavior follows directly from  the classical solution via Eq. (\ref{tauvariance}): the coherent large-angle memory of each Planck-energy single-particle state is independent of $R$, so the total distortion grows with the  duration in the same way as a random walk.

Our semiclassical treatment meshes well with previous arguments that suggest macroscopic coherence of geometrical quantum states.
For physical
field states  not to exceed the black hole mass in any volume,  gravity  
 requires macroscopic nonlocal  coherence in the infrared\cite{CohenKaplanNelson1999,PhysRevD.101.126010,2021arXiv210304509C}.
 Nonlocally  coherent  geometrical states (``entanglement wedges'')  are also key elements in recent resolutions of the black hole information paradoxes, incorporated into an account of fine-grained entropy
\cite{Almheiri:2020cfm}.
Formal methods based on conformal descriptions of near-horizon states show that  coherence can produce physical metric distortions  comparable in amplitude to those estimated here,  both for black hole horizons and  causal diamond surfaces in  flat space-time\cite{Zurek:2020ukz,Banks:2021jwj}. 

The time distortions on the causal diamond surface can be continuously extrapolated to provide a  physical heuristic picture of what happens when a black hole evaporates. As a black hole evaporates (or is assembled) one particle at a time, each particle maps onto a coherent, mostly quadrupolar distortion that extends across  the entire horizon surface, of the order of one Planck length in amplitude.  A long-lived horizon is a  superposition of a sequence of many such distortions. This simple  picture of coherent horizon distortions  agrees with a  ``quantum-first'' analysis of evaporation\cite{Giddings_2017,Giddings:2019vvj}. In this view, states of Planck-scale ``quantum foam'' are wave functions coherently spread everywhere across the horizon, no matter how large it is.
  
\section{Conclusion}

The Schr\"odinger-cat-like states  of  space-times with quantum decaying particles---   superpositions of different geometries with macroscopically distinguishable physical properties, such as quadrupole moments of observable time distortions---   show  concretely how  quantum-gravitational coherence works. The quantum construction uses particle-like   states of matter, with superpositions of wave functions localized to causal diamonds, instead of quantized plane-wave states that extend to infinity, so it  allows both preparation and measurement of states from a single world line.
Directional indeterminacy of  quantum states is reconciled with gravity via a conventional correspondence principle, leading to quantum-geometrical states that are directionally coherent on null surfaces.
 It explicitly shows  nonlocal, anisotropic effects of gravitational coherence manifested in a local measurement.


Our analysis shows that macroscopic coherence of quantum gravity  has concrete physical consequences. 
Coherence leads to  quantum-gravitational distortions on large angles and macroscopic scales, with variance  that grows linearly with duration.
The smaller variation  predicted by   effective field theory does not account for   the physical effects of directional causal coherence.

The   anisotropic distortion of  time is a real physical effect,  not a gauge artifact: in principle, it can be measured with actual clocks or laser interferometers. The permanent time displacement  between the observer's clock and any of the other clocks is an objective measurable quantity. Different states correspond to physically different geometries on macroscopic scales.

It is possible that the coherence of active quantum gravity could  be measured not just in principle, but in an actual experiment. Although the distortions of physical black hole horizons are likely to be unobservable,
  a direct measurement of causal diamond distortions may be accessible to Michelson interferometers correlated to reveal spacelike coherence\cite{holoshear,PhysRevLett.126.241301}. The effect of coherent horizons  would
  produce primordial perturbations during cosmic inflation of the order of Eq. (\ref{Deltamag}), much larger than standard inflation theory, and could
  create new causally-coherent symmetries of angular correlations in cosmic microwave background anisotropy. These symmetries appear to be consistent with anomalous correlations measured in the CMB on large angular scales\cite{PhysRevD.99.063531,Hogan_2020,Hagimoto_2020,hogan2021angular}.



 \begin{acknowledgments}
 We are grateful for useful discussions with R. Wald.
 This work was supported by the Department
 of Energy at Fermilab under Contract No. DE-AC02-07CH11359.  
\end{acknowledgments}

\bibliography{CCbib}

\section{Appendix}
\subsection{Tensor Spherical Harmonics}
In this section we lay out some properties of tensor spherical harmonics that will be used for the mode sum in section III. Consider a field with spin $s$. We can decompose any tensor on the sphere into a mode sum over rank $s$ tensors weighted by spherical harmonics $Y_{l,m}$ in the following manner
\begin{align}
    \mathcal{Y}_{j,l,s,m} &= \sum_{m_l=-l}^l \sum_{m_s=-s}^s \langle l,m_l;s,m_s|j, m\rangle Y_{l,m_l} \tilde{t}_{s, m_s} \\
    &= \sum_{m_s=-s}^s \langle l,m-m_s;s,m_s|j, m\rangle Y_{l,m-m_s} \tilde{t}_{s, m_s} 
\end{align}
where the $\tilde{t}_{s,m_s}$ are a basis of spin $s$ unit tensors satisfying
\begin{equation}
    \tilde{t}^*_{\alpha} \cdotp \tilde{t}_{\beta}=\delta_{\alpha,\beta}
\end{equation}
and the $\langle \_|\_\rangle$ are Clebsch-Gordon coefficients. For the case of gravitational radiation, we will be concerned with $s=2$ tensor harmonics, specifically the transverse ones, which can be separated into ``electric'' (E) and ``magnetic'' (M) parts. Here, the dot product between spin 2 tensors is taken to mean
\begin{equation}
    A\cdotp B  = \sum_{i,j} A_{i,j}B_{i,j}
\end{equation}

\subsection{Transverse Tensor Spherical Harmonics}
As mentioned above, there are two types of transverse tensor harmonics. 
\begin{align}
    \mathcal{Y}^E_{l,m} &= \sqrt{\frac{(l+1)(l+2)}{2(2l+1)(2l-1)}} \mathcal{Y}_{l, l-2, m}+\sqrt{\frac{3(l-1)(l+2)}{(2l-1)(2l+3)}} \mathcal{Y}_{l, l, m} \\ \nonumber
    &+\sqrt{\frac{l(l-1)}{2(2l+1)(2l+3)}} \mathcal{Y}_{l, l+2, m} \\
    \mathcal{Y}^M_{lm} &= \sqrt{\frac{l+2}{2l+1}} \mathcal{Y}_{l, l-1, m}+\sqrt{\frac{l-1}{2l+1}} \mathcal{Y}_{l, l+1, m}
\end{align}
These satisfy orthogonality conditions given by
\begin{align}
    \int d\Omega \mathcal{Y}^E_{lm} \cdotp \mathcal{Y}^{E*}_{l'm'} = \delta_{l l'}\delta_{m m'} \\
    \int d\Omega \mathcal{Y}^M_{lm} \cdotp \mathcal{Y}^{M*}_{l'm'} = \delta_{l l'}\delta_{m m'}
\end{align}
One can show using a series of recursion relations that the transverse harmonics have the following explicit form
\begin{align}
    \mathcal{Y}^E_{l,m} &= (\Hat{\theta} \Hat{\theta} - \Hat{\phi}\Hat{\phi}) \frac{1}{\sqrt{2l(l+1)(l-1)(l+2)}}\left(2\frac{\partial^2}{\partial \theta^2}+l(l+1)\right)Y_{l,m} \\ \nonumber
    &+(\Hat{\theta} \Hat{\phi} + \Hat{\phi} \Hat{\theta})\sqrt{\frac{2}{l(l+1)(l-1)(l+2)}} i m \frac{\partial}{\partial \theta}\left(\frac{Y_{l,m}}{{\rm sin}\theta}\right) \\
    \mathcal{Y}^M_{lm} &=- (\Hat{\theta} \Hat{\theta} - \Hat{\phi}\Hat{\phi}) \sqrt{\frac{2}{l(l+1)(l-1)(l+2)}}m \frac{\partial}{\partial \theta}\left(\frac{Y_{l,m}}{{\rm sin}\theta}\right) \\ \nonumber
    &-(\Hat{\theta} \Hat{\phi} + \Hat{\phi} \Hat{\theta})\frac{1}{\sqrt{2l(l+1)(l-1)(l+2)}}\left(2\frac{\partial^2}{\partial \theta^2}+l(l+1)\right)Y_{l,m}
\end{align}
\subsection{Memory Decomposition}
Now we will take the result from \cref{dispkick} and expand in a basis of transverse tensor harmonics. Note importantly that the transverse tensor harmonics are not defined for $l<2$. In the case we are considering, we have axisymmetry, so only the $m=0$ modes contribute. Looking at the form of \cref{dispkick}, we see that we must have that 
\begin{equation}
    \Delta D = \frac{M}{r} \left(\sum_{l=2}^{\infty} a_l \mathcal{Y}^E_{l,0} \right)\cdotp D_0
\end{equation}
We can determine the $a_l$ using the orthogonality condition eq. (60). 

\begin{align}
    a_l &= \int d\Omega \mathcal{Y}^{E*}_{l,0} \cdotp (\Hat{\theta}\Hat{\theta}-\Hat{\phi}\Hat{\phi}) \nonumber \\
    &=\sqrt{\frac{2}{l(l+1)(l-1)(l+2)}}\sqrt{\frac{2l+1}{4\pi}} \int d\Omega \left(2\frac{d^2}{d \theta^2}+l(l+1)\right)P_l({\rm cos}\theta) \nonumber \\
    &= -\sqrt{\frac{2\pi(2l+1)}{l(l+1)(l-1)(l+2)}} \int d{\rm cos}\theta \left(2{\rm cot}\theta\frac{d}{d \theta}+l(l+1)\right)P_l({\rm cos}\theta) \nonumber \\
    &= 2\sqrt{\frac{2\pi(2l+1)}{l(l+1)(l-1)(l+2)}} \int dx x \frac{d}{dx}P_l(x) \nonumber \\
    & =  4\sqrt{\frac{2\pi(2l+1)}{l(l+1)(l-1)(l+2)}} \quad (l \: {\rm even}) \label{tecoeff}
\end{align}

\end{document}